\documentclass[aps,prb,twocolumn]{revtex4-2}

\usepackage{amsmath,amssymb,graphicx}
\usepackage{tikz}
\usepackage{graphicx}
\usepackage{array}
\usepackage[margin=1in]{geometry}
\usepackage{color}
\usepackage{array}
\usepackage{booktabs}
\usepackage{float}
\usepackage{hyperref}
\hypersetup{
    colorlinks=true,
    linkcolor=magenta,
    citecolor=magenta,
    urlcolor=magenta
}
\newcommand\bea{\begin{eqnarray}}
\newcommand\eea{\end{eqnarray}}
\newcommand\beq{\begin{equation}}
\newcommand\eeq{\end{equation}}
\def\nn{\nonumber}
\def\f{\frac}
\def\al{\alpha}

\def\si{\sigma}

\def\th{\theta}

\begin{document}

\title{ Electrically tunable Crossed Andreev Reflection in Bipolar Magnetic Semiconductors}

\author{Polireddi Naveen}
\author{Abhiram Soori}
\email{abhirams@uohyd.ac.in}
\affiliation{School of Physics, University of Hyderabad, Prof. C. R. Rao Road, Gachibowli, Hyderabad-500046, India}
\begin{abstract}
Crossed Andreev reflection (CAR) is a nonlocal quantum transport phenomenon that arises at the interfaces between a superconductor  and two spatially separated metals. In this process, an electron incident from one metal combines with another electron originating from the other metal to form a Cooper pair in the SC. In this work, we theoretically investigate the quantum transport properties of a device comprising a central $s$-wave superconductor sandwiched between two bipolar magnetic semiconductor (BMS) leads. BMS materials possess a unique bandstructure where the conduction and valence bands are fully spin-polarized in opposite directions. By applying independent gate voltages to the two BMS leads, we selectively tune their chemical potentials to configure an antiparallel spin alignment at the Fermi energy.
In the absence of spin-orbit coupling (SOC), this specific configuration perfectly suppresses both local Andreev reflection and direct electron tunneling (ET), thereby isolating and profoundly enhancing CAR. The introduction of Rashba SOC, however, induces spin-flip scattering that relaxes these strict selection constraints, yielding finite probabilities for competing scattering channels and subsequently reducing the overall CAR efficiency. Nevertheless, for SOC strengths below a critical threshold, CAR remains the dominant transport mechanism over ET.
This proposed architecture provides a scalable, stray-field-free method to achieve purely electrostatic control over nonlocal transport, advancing the development of highly efficient superconducting spintronic devices.
\end{abstract}

\maketitle
\section{Introduction}

The burgeoning field of superconducting spintronics seeks to exploit the electron spin degree of freedom to facilitate low-power, high-speed information processing, a pursuit increasingly vital for quantum device architectures. In this field, crossed Andreev reflection (CAR) has emerged as a fundamental mechanism for the generation and manipulation of nonlocal, spin-entangled electron pairs~\cite{Yeyati2007,PhysRevB.63.165314,PhysRevLett.104.026801,Hofstetter}. Typically realized in a device geometry where two metallic leads are coupled to a central   SC at distinct locations, CAR occurs when an electron incident from one normal lead is reflected as a hole in the second, spatially separated lead. The inverse of this process corresponds to the splitting of a Cooper pair from the SC into two spatially distinct metals, resulting in nonlocally entangled electrons. Consequently, this nonlocal transport process provides a robust pathway for realizing entangled electronic states in solid-state systems.

Bipolar magnetic semiconductors (BMSs) constitute a unique class of materials characterized by a band structure in which the conduction and valence bands are fully spin-polarized in opposite directions~\cite{C2NR31743E,PhysRevMaterials.5.034005,LI2022511,Chen2023Recent,Pang2020Bipolar,Deng2022}. This unique electronic configuration enables precise electrical control over spin orientation: by tuning the gate voltage to shift the Fermi level between these bands, the carrier type and spin polarization can be reversibly switched between fully spin-up and fully spin-down states~\cite{C3CP52623B,Li2023}. In contrast, while half-metals can achieve nearly 100\% spin polarization~\cite{Son2006,PhysRevB.110.085130,5z4v-x6xn}, they typically support only a single spin channel. Reversing the spin orientation in half-metallic systems generally necessitates flipping the magnetization direction via external magnetic fields, a process that is inherently slower and more energy-intensive than electrostatic gating.

In practical device geometries, nonlocal transport is often dominated by electron tunneling (ET)---the direct tunneling of electrons between leads across the SC---which tends to mask the signature of CAR in conductance measurements. Conventional strategies to isolate or enhance CAR involve utilizing ferromagnetic(FM) elements and  spin-selective transport~\cite{PhysRevLett.93.197003,PhysRevB.70.174509,PhysRevB.72.165322,SOORI2022114721,Melin2005,Giazotto2006,Kalenkov2007}. However, the reliance on macroscopic magnetic elements introduces stray fringing fields that can interfere with neighboring quantum components and complicate experimental scalability. To mitigate these issues, alternative platforms such as altermagnets(AM) have been proposed, as they exhibit spin-split electronic structures without a net macroscopic magnetization, thereby eliminating stray fields while retaining control over spin-dependent transport~\cite{Das2024,wang2026,Niu2024}.
Carbon nanotube-based systems to enhance CAR can benefit from strong Coulomb interactions; however, competing ET and stringent spin--valley selection rules may limit CAR efficiency~\cite{Burset2011}. 
Alternatively, two-dimensional honeycomb antiferromagnets, such as graphene-based systems and silicene, offer a tunable platform for spin-dependent superconducting transport through electrostatic control~\cite{Jakobsen2021,Li2024}. However, transport is highly sensitive to system geometry; macroscopic misalignment of Néel vectors can suppress non-local CAR in favor of ET~\cite{Jakobsen2021}, while atomic-scale interfacial scattering governs valley-mixing in local Andreev reflection~\cite{Li2024}.

\begin{table*}[t]
\centering
\renewcommand{\arraystretch}{2.5}
\begin{tabular}{|c|c|c|}
\hline
\textbf{CAR systems} & \textbf{Pros} & \textbf{Cons} \\
\hline

FM-based systems~\cite{PhysRevLett.93.197003,PhysRevB.70.174509,PhysRevB.72.165322,SOORI2022114721,Melin2005,Giazotto2006,Kalenkov2007} &
\parbox{4cm}{\renewcommand{\baselinestretch}{1.2}\selectfont
Experimentally realized} &
\parbox{6cm}{\renewcommand{\baselinestretch}{1.2}\selectfont
\vspace{0.1cm}
Require external magnetic fields, which can generate unwanted
stray magnetic fields.}\vspace{0.05cm} \\
\hline
\vspace{0.02cm}
AM-based systems~\cite{Das2024,wang2026,Niu2024} &
\parbox{4cm}{\renewcommand{\baselinestretch}{1.2}\selectfont
Do not require an external magnetic field} &
\parbox{6cm}{\renewcommand{\baselinestretch}{1.2}\selectfont
Require knowledge of the crystallographic orientation.} \\
\hline
BMS-based systems &
\parbox{4cm}{\renewcommand{\baselinestretch}{1.2}\selectfont
Eliminate the need for an external magnetic field} &
\parbox{6cm}{\renewcommand{\baselinestretch}{1.2}\selectfont
SOC suppresses the CAR probability.} \\
\hline
\end{tabular}
\caption{Comparison of different CAR systems, highlighting their pros and cons.}
\label{tab:example}
\end{table*}

\begin{figure}[ht]
\includegraphics[width=8cm]{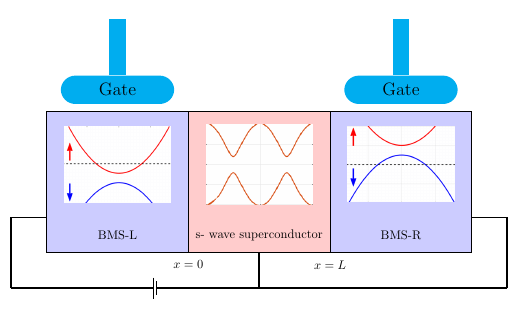}
    \caption{Schematic illustration of the proposed device geometry, comprising a central $s$-wave SC  sandwiched between two BMS leads in the absence of SOC. Independent, tunable gate voltages applied to each BMS electrode provide local control over the chemical potential. By tuning the respective Fermi energies such that the charge carriers participating in transport in the left and right leads possess opposite spin polarizations, crossed Andreev reflection   is significantly enhanced while competing electron tunneling is naturally suppressed.}\label{fig:schem}
    \end{figure}

In this work, we propose a device geometry comprising a central SC coupled to two BMS leads as shown in Fig.~\ref{fig:schem}. By independently tuning the chemical potentials of the BMS electrodes, we can configure the system such that the charge carriers at the Fermi energy in the respective leads possess opposite spin polarizations. This gate-defined antiparallel spin alignment strongly enhances CAR while naturally suppressing competing ET. Crucially, this approach enables purely electrical tuning of CAR without the necessity of external magnetic fields. Such electrostatic control significantly simplifies the device architecture and enhances the potential for scalable quantum integration, reflecting a broader shift in quantum transport engineering away from magnetically driven configurations and toward highly efficient, electrically tunable mechanisms.

\section{Methods for BMS-SC-BMS junction}
We consider a two-dimensional BMS described by the Hamiltonian
\[
H_{\vec k} = \Big(t_bk^2a^2+b\Big)\si_z\tau_z+\delta k^2a^2\tau_z-\mu_{bs}\tau_z+H_{SOC}, 
\]
 where  $H_{SOC}=\alpha(k_y\si_x-k_x\si_y\tau_z)$ is Rashba spin-orbit coupling term, 
$t_b = \frac{\hbar^2}{2ma^2}$, $m$ is the effective mass of the BMS bands in the limit of equal mass magnitudes, and the parameter $\delta$ ($0 < \delta < t_b$), which carries the same units as $t_b$, characterizes the difference between the magnitudes of the two effective masses,
 $b$ characterizes the bandgap of the semiconductor separating the up-spin (conduction) and  down-spin (valence)  bands, $m$ is the effective mass,  \(\mu_{b}\) is the chemical potential,  \(\sigma _{j}\) are the  Pauli spin matrices acting on the spin-space,  $\tau_j$ are the Pauli spin matrices acting on the particle-hole space, and \(\vec k = (k_{x},k_{y})\) is the wavevector for electrons.
A singlet-superconductor is sandwiched between two BMS leads. 
The superconducting Hamiltonian mixes electron and hole states with opposite spin projections, i.e., $\uparrow e \leftrightarrow \downarrow h$ and $\downarrow e \leftrightarrow \uparrow h$. 
We calculate local conductivity \(G_{LL} = {dI_{L}}/{dV}\) and nonlocal conductivity \(G_{RL} = {dI_{R}}/{dV}\) where \(I_{L}\) and \(I_{R} \) are the current densities  in the left and right BMS, $V$ is the voltage bias applied from left BMS, keeping SC and the right BMS grounded. We use  Landauer-Büttiker scattering formalism~\cite{soori17}.

\begin{figure}
\includegraphics[width=7cm]{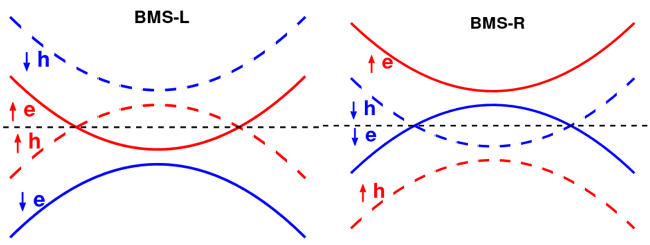} 
\caption{Dispersion relations of the left and right BMS  for $\alpha = 0$ ,$\mu_{bl} = -\mu_{br} = 4b$. The red (blue) lines represent the up-spin (down-spin) electron channels, while the dotted lines correspond to the respective  hole channels.}\label{fig:bms-disp}
\end{figure}

We choose the chemical potential on the left $\mu_{bl}>0$ and that on the right $\mu_{br}<0$. This ensures that in the limit of $\alpha=0$, near the Fermi energy up-spin electrons  are present in the left BMS whereas down-spin electrons  are present in the right BMS. The Hamiltonian for the full system is given by 
\begin{widetext}
\begin{equation}
H =
\begin{cases}
-(t_b\sigma_z\tau_z + \delta\tau_z)a^2\vec\nabla^2 +b\sigma_z\tau_z - \mu_{bl}\tau_z -i\alpha(\sigma_x\frac{\partial}{\partial y}-\sigma_y\tau_z\frac{\partial}{\partial x})& \text{ for  }x < 0 \\ 
-t_s\tau_za^2\vec\nabla^2 -\mu_s\tau_z +\Delta \si_y\tau_y
 & \text{ for } 0<x<L \\
-(t_b\sigma_z\tau_z + \delta\tau_z)a^2\vec\nabla^2 +b\sigma_z\tau_z - \mu_{br}\tau_z -i\alpha(\sigma_x\frac{\partial}{\partial y}-\sigma_y\tau_z\frac{\partial}{\partial x})& \text{ for } x>L
\end{cases}
\end{equation}
\end{widetext}
 where  $\Delta$ denotes the superconducting pairing strength and $\mu_{s} $ denotes the chemical potential of the superconductor. 
 In the presence of SOC, the dispersion relations for the BMS regions are given by
\begin{align}
    E &= \delta k^2a^2 - \mu_{bs} \pm \sqrt{\al^2k^2 + (b + t_bk^2a^2)^2}, \\
    E &= -\delta k^2a^2 + \mu_{bs} \pm \sqrt{\al^2k^2 + (b + t_bk^2a^2)^2},
\end{align}
where the former and latter equations correspond to electrons and holes, respectively, with $k^2 = k_x^2 + k_y^2$. The chemical potential $\mu_{bs}$ is chosen as $\mu_{bl}$ ($\mu_{br}$) for the left (right) BMS region. The dispersion relations on both sides are shown in Fig.~\ref{fig:bms-disp} in the limit $\alpha = 0$, where $\sigma_z$ is a good quantum number.

  The dispersion relation  in the superconductor is 
\[ E  =
\pm \sqrt{\left[\frac{\hbar^2(q_x^2 + q_y^2)}{2m} - \mu_s\right]^2 + \Delta^2}
\]

The eigenstates  in superconductor are Bogoliubov–de Gennes (BdG) quasiparticles, which have both the components: electron and hole. In the SC region, when the energy is within the superconducting energy gap, the BdG  states are evanescent modes and are equally electron like and hole like.

For $\delta, b > 0$, the electron band in the left BMS region that curves upward at large $k$ has a minimum at energy $|b| - \mu_{bl}$. Above this threshold energy, an electron can be incident from the left BMS region. The scattering eigenfunction corresponding to an electron incident at energy $E > |b| - \mu_{bl}$ with an angle of incidence $\theta$ takes the form $\psi e^{i k_y y}$, where
\begin{widetext} 
\begin{equation}
\psi(x)=
\begin{cases}
\frac{e^{ik_{xe1l}x}}{N_1^+}\begin{pmatrix}u_{e1l}^+\\v_{e1l}\\0\\0
\end{pmatrix}
+r_{e1}e^{-ik_{xe1l}x}\begin{pmatrix}u_{e1l}^-\\v_{e1l}\\0\\0
\end{pmatrix}
+r_{e2}e^{-ik_{xe2l}x}\begin{pmatrix}u_{e2l}^-\\v_{e2l}\\0\\0
\end{pmatrix} \\
~~~+r_{h1}e^{ik_{xh1l}x}\begin{pmatrix}0\\0\\u_{h1l}\\v_{h1l}
\end{pmatrix}
+r_{h2}e^{ik_{xh2l}x}\begin{pmatrix}0\\0\\u_{h2l}\\v_{h2l}
\end{pmatrix},
& x < 0 \\ 
\displaystyle \sum_{j=1}^{4} A_j e^{iq_{xj}x}\begin{pmatrix}
    u_j \\ 0 \\ 0\\ -\Delta \end{pmatrix} + \displaystyle \sum_{j=1}^{4}B_j e^{iq_{xj}x} \begin{pmatrix}
    0 \\ u_j \\ \Delta\\ 0 \end{pmatrix},
    & 0<x<L \\

t_{e1}e^{ik_{xe1r}(x-L)}\begin{pmatrix}u_{e1r}^+\\v_{e1r}\\0\\0
\end{pmatrix}

+t_{e2}e^{ik_{xe2r}(x-L)}\begin{pmatrix}u_{e2r}^+\\v_{e2r}\\0\\0
\end{pmatrix} \\ 
~~~+t_{h1}e^{ik_{xh1r}(x-L)}\begin{pmatrix}0\\0\\u_{h1r}\\v_{h1r}
\end{pmatrix}
+t_{h2}e^{ik_{xh2r}(x-L)}\begin{pmatrix}0\\0\\u_{h2r}\\v_{h2r}
\end{pmatrix},
& x > L \\ 
\end{cases}
\end{equation}
\end{widetext} 
Here, 
\begin{align}
    k_{e1l} &= \sqrt{\frac{-\eta_e + \sqrt{\eta_e^2 - 4(t_b^2-\delta^2)a^4\big(b^2-(E+\mu_{bl})^2\big)}}{2(t_b^2-\delta^2)a^4}}, \\
    \eta_e &= 2bt_ba^2 + \alpha^2 + 2\delta a^2(E+\mu_{bs}), ~~~\nn \\ k_y&=k_{e1l}\sin{\theta},~~k_{xe1l}=k_{e1l}\cos\th
\end{align}
The other wavenumbers in the BMS regions, denoted by $k_{pjsx}$, are determined from their respective dispersion relations to ensure the wave is either outgoing from the junction or evanescent. Here, the indices denote the particle type ($p = e, h$ for electron or hole), the specific root for $k_{pjsx}$ from the dispersion relation ($j = 1, 2$), and the junction side ($s = l, r$ for left or right).

\begin{equation}
\begin{aligned}
u_{ejs}^{\pm} &= \alpha\left(\pm i k_{xejs} + k_y\right), \\
v_{ejs} &= E + \mu_{bs} - b - (t_b+\delta)k_{ejs}^{2}a^2,
\qquad j=1,2, \\
k_{ejs}^{2}&=k_{xejs}^2+k_y^2,~~N_1^+ = \sqrt{|u_{e1l}^+|^2+|v_{e1l}|^2}, \\
u_{hjs} &= \alpha\left( i k_{xhjs} + k_y\right), ~~k_{hjs}^2=k_{xhjs}^2+k_y^2,\\
v_{hjs} &= E - \mu_{bs} + b + (t_b+\delta)k_{hjs}^{2}a^2,
\qquad j=1,2, \\[6pt]
u_j &= E + \xi_j, \\
\xi_j &=  t_sa^2(q_{xj}^2+k_y^2)-\mu_{s},
\qquad j=1,2,3,4.
\end{aligned}
\end{equation}

\begin{align}
q_{x1} &=
\ \sqrt{\frac{\mu_s + \sqrt{E^2 - \Delta^2}}{t_sa^2}
 - k_{y}^2} 
\ , q_{x2} = -q_{x1} \nn \\
q_{x3} &=
\ \sqrt{\frac{\mu_s - \sqrt{E^2 - \Delta^2}}{t_sa^2}
 - k_{y}^2} 
\ , q_{x4} = -q_{x3} 
\end{align}

The scattering coefficients $r_{e1},\,r_{e2},\,r_{h1},\,r_{h2},$ $t_{e1},\,t_{e2},\,t_{h1},\,t_{h2},\,A_j,\,B_j$ are determined by applying the following boundary conditions, which are derived by enforcing the conservation of probability current.
\bea
\psi(0^-) &=& \psi(0^+), \nn \\
\Big(t_sa^2\frac{\partial \psi(0^+)}{\partial x} && \nn \\ -a^2(t_b\sigma_z + \delta\sigma_0)\frac{\partial  \psi(0^-)}{\partial x}+\frac{i\alpha}{2}\sigma_y\psi(0^-)\Big)
&=&t_sq_0\psi(0), \nn \\
\psi(L^-) &=& \psi(L^+), \nn \\
\Big(a^2(t_b\sigma_z + \delta\sigma_0)\frac{\partial  \psi(L^+)}{\partial x}-\frac{i\alpha}{2}\sigma_y\psi(L^+)&& \nn \\ - t_sa^2\frac{\partial \psi(L^-)}{\partial x}\Big)
&=&t_sq_0\psi(L), \nn \\ 
\eea
Here, the parameter $q_0$ corresponds to the strength of delta function barrier at the interface. 

The expression for the probability current density in the BMS region is given by 
\begin{equation}
    J_P = (2/\hbar){\rm Im}\Big(\psi^\dagger(t_b\sigma_z+\delta\sigma_0)\tau_za^2\frac{\partial\psi}{\partial x} - i\psi^\dagger\frac{\alpha}{2}\sigma_y\tau_z\psi\Big)
\label{eq:Jp}
\end{equation}
The expression for the charge current density is given by 
\begin{equation}
    J_c = (2e/\hbar){\rm Im}\Big(\psi^\dagger(t_b\sigma_z+\delta\sigma_0)\tau_0a^2\frac{\partial\psi}{\partial x} - i\psi^\dagger\frac{\alpha}{2}\sigma_y\tau_0\psi\Big)
\label{eq:Jc}
\end{equation}
The local and nonlocal differential conductivities ($G_{LL}$ and $G_{RL}$) are given by 
\bea 
G_{LL} &=& \f{e^2}{4\pi^2} \Gamma(E) \int_{-\pi/2}^{\pi/2}J_{cl}(E,\th)d\th, \nn \\ 
G_{RL} &=& \f{e^2}{4\pi^2} \Gamma(E) \int_{-\pi/2}^{\pi/2}J_{cr}(E,\th)d\th, \nn  \\
 \Gamma(E)&=&\Big[{2\delta + \frac{2t_ba^2(t_bk_{e1l}^2a^2+b)+\alpha^2}{\sqrt{(t_bk_{e1l}^2a^2+b)^2+\alpha^2k_{e1l}^2}}}\Big]^{-1}
\label{eq:conductivity}\eea
where $J_{cl}$ ($J_{cr}$) is charge current density evaluated in the region $x<0$ ($x>L$).

\begin{figure*}[htbp]
    \centering

    \includegraphics[width=0.48\textwidth]{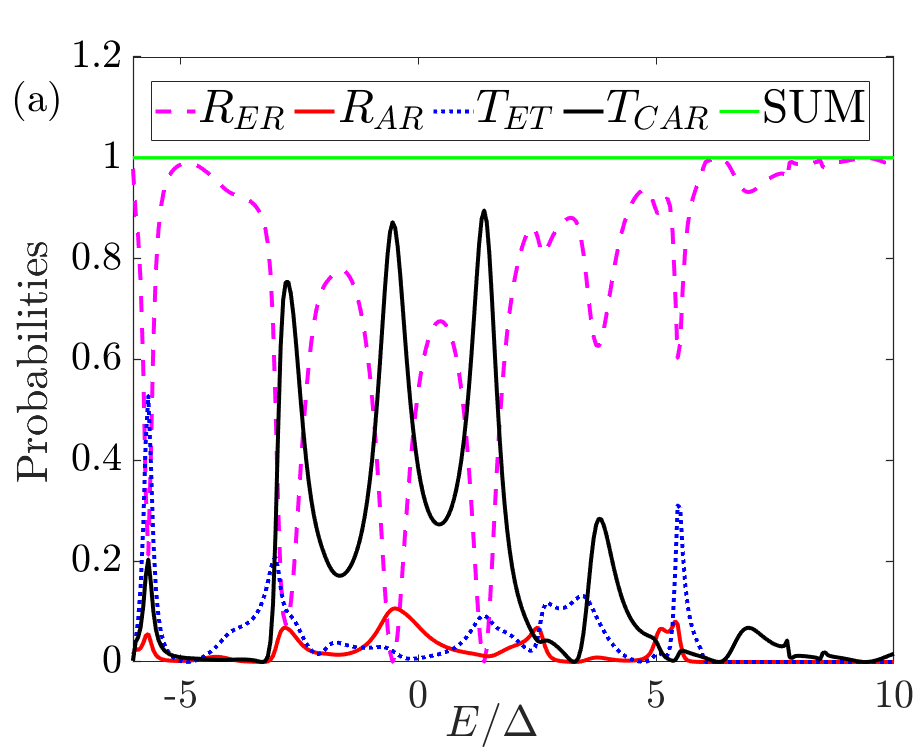}
    \hfill
    \includegraphics[width=0.48\textwidth]{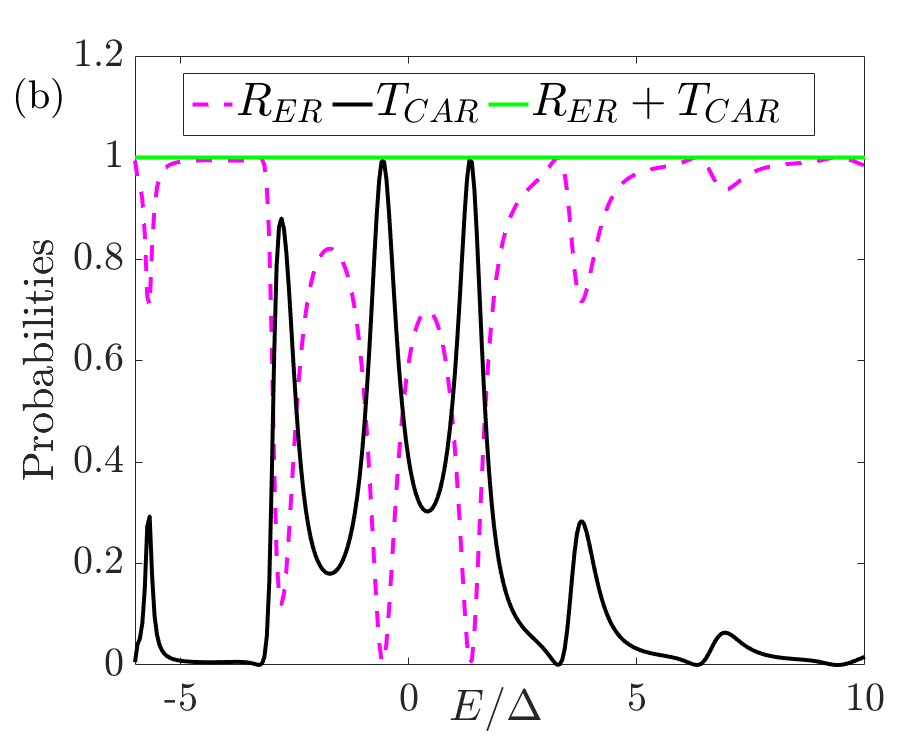}

    \vspace{0.3cm}

    \includegraphics[width=0.48\textwidth]{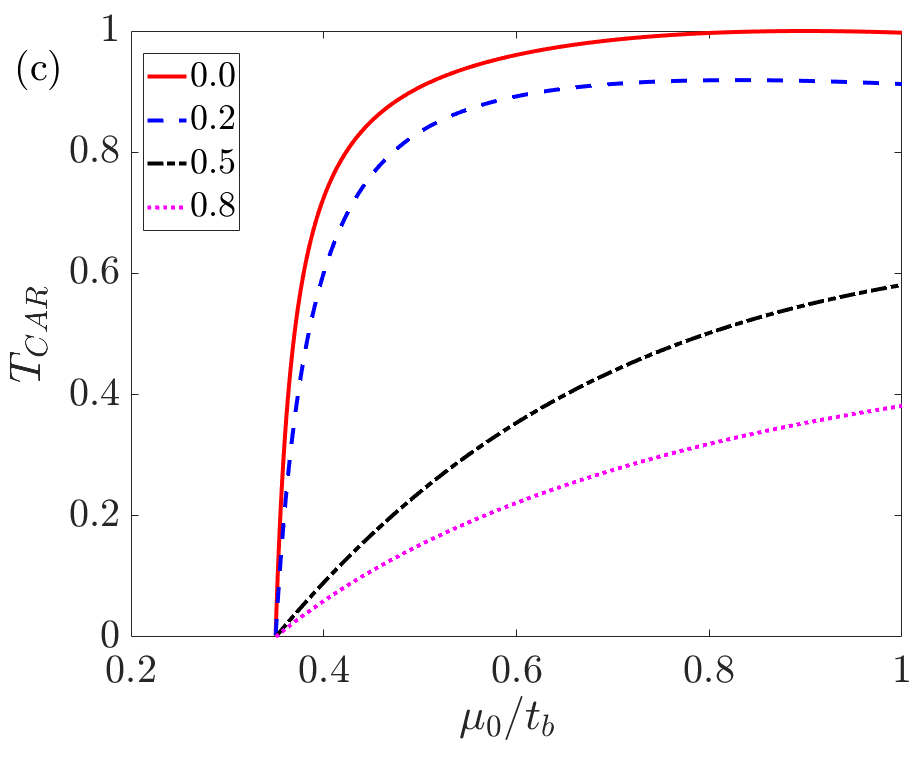}
    \hfill
    \includegraphics[width=0.48\textwidth]{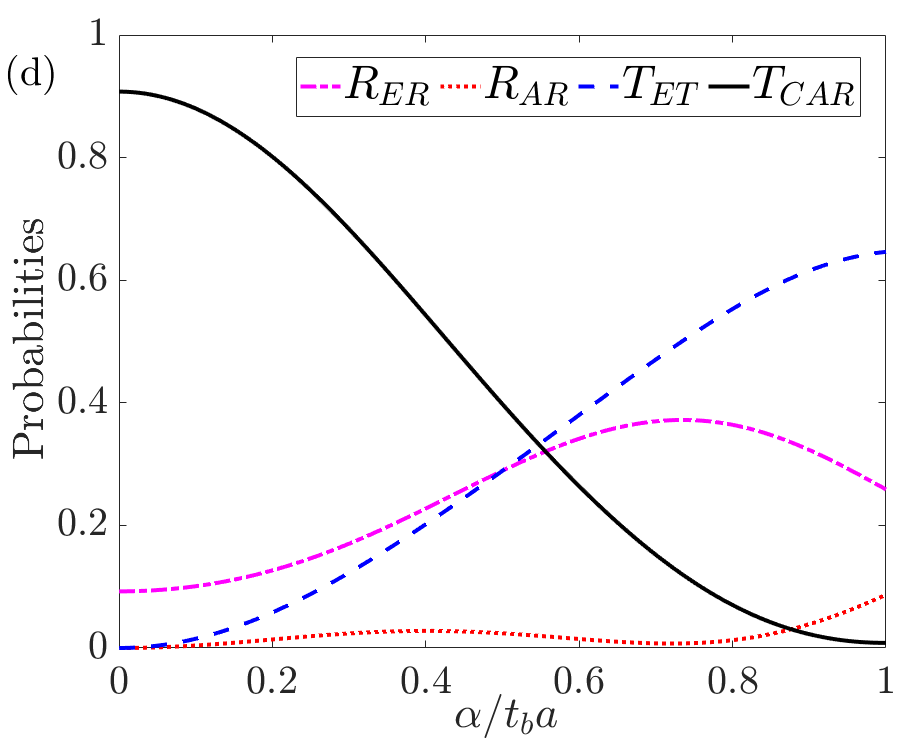}

    \caption{The probabilities of the scattering processes  (a) versus the incident energy in the presence of SOC ($\alpha=0.2t_b$) in BMS, (b) Versus the incident energy neglecting SOC in BMS (c)$T_{CAR}$ versus chemical potential $\mu_0/t_b$ for different values of $q_0$ in absence of SOC at $E = 1.5\Delta$, (d) probabilities versus $\alpha/t_ba$ at $E = 1.5\Delta$ and for  $\mu_{bl}=-\mu_{br}=0.5t_b$,  $\mu_s=50\Delta=t_b$, for $\theta=0$, $\Delta=0.02t_b$, $b=0.38t_b$, $L=100.6a$, and $q_0=0.2\sqrt{\mu_s/t_sa^2}$.}
    \label{fig:probabilities_SOC}
\end{figure*}

\section{Results and discussion}
To ensure the physical relevance of our model, we adopt parameters representative of realistic material systems. For the BMS leads, we base our values on $VNbRuAl$  , an experimentally realized bipolar magnetic semiconductor that features a 0.65 eV energy gap between its oppositely spin-polarized conduction and valence bands~\cite{PhysRevB.104.134406}. For the central superconducting region, we selected $La_2PrNiO_7$, which exhibits a superconducting gap of 19~meV and  s-wave pairing symmetry~\cite{SC_expmaterial}. We select the chemical potentials $\mu_{bl} = -\mu_{br} = 0.5t_b$, along with $\mu_s = 50\Delta$, $b = 0.384t_b$, $m_s = 0.11m_e$, $\delta/t_b = 0.1$ and the lattice constant $a = 6.24\AA$, where $m_e$ is the bare electron mass and $\alpha = 0.2t_ba$~\cite{SOC} . 

Our  objective is to investigate the influence of  SOC on these competing scattering processes, with particular emphasis on determining whether SOC suppresses the efficiency of  CAR. In the absence of SOC, when the chemical potentials of the two BMS leads are tuned to opposite signs, the system exhibits a favorable regime for CAR, in which both local AR and ET are completely suppressed. This suppression arises from the unavailability of the down-spin hole band on the left BMS and the up-spin electron band in the right BMS as shown in Fig.~\ref{fig:bms-disp}. Consequently, the transport regime becomes completely dominated by direct competition between ER and CAR. Under these conditions, the incident electron can undergo CAR  with enhanced probability, allowing for efficient nonlocal electron--hole conversion between the two leads.  Depending on the form and strength of the SOC, this coupling can induce spin-flip scattering processes, thereby relaxing the spin-selection constraints that suppress certain scattering channels in the absence of SOC. Consequently, processes that are otherwise forbidden or strongly suppressed, particularly AR and ET, may acquire finite probabilities. The emergence of these additional scattering channels redistributes the total scattering probabilities among the competing processes and can reduce the probability associated with CAR.

\begin{figure*}[htbp]
\centering

\begin{minipage}[t]{0.48\textwidth}
    \centering
    \includegraphics[width=\linewidth]{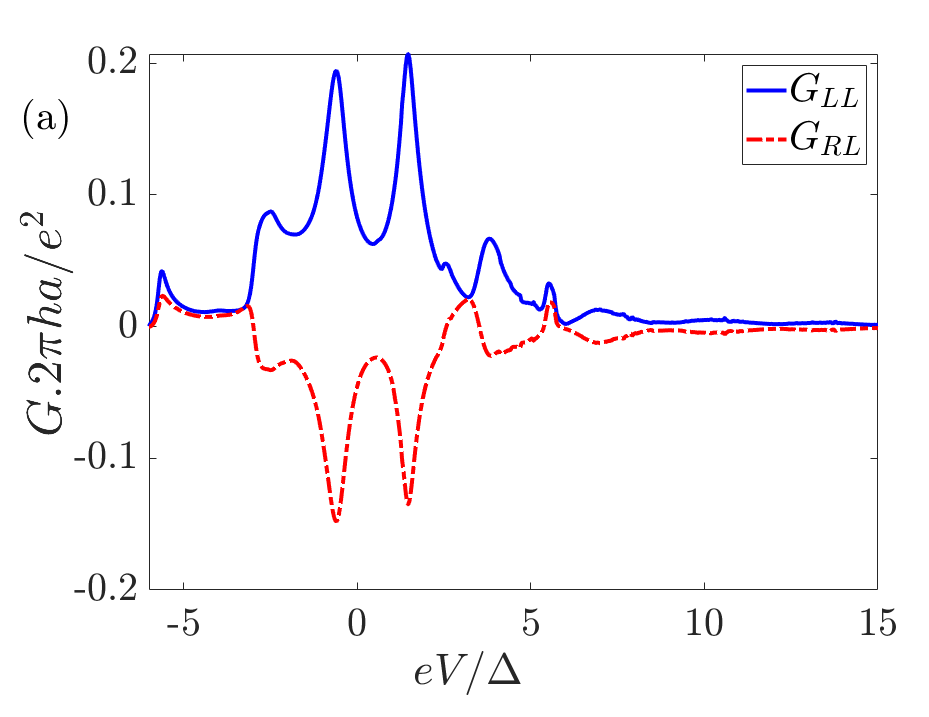}
    
\end{minipage}
\hfill
\begin{minipage}[t]{0.48\textwidth}
    \centering
    \includegraphics[width=\linewidth]{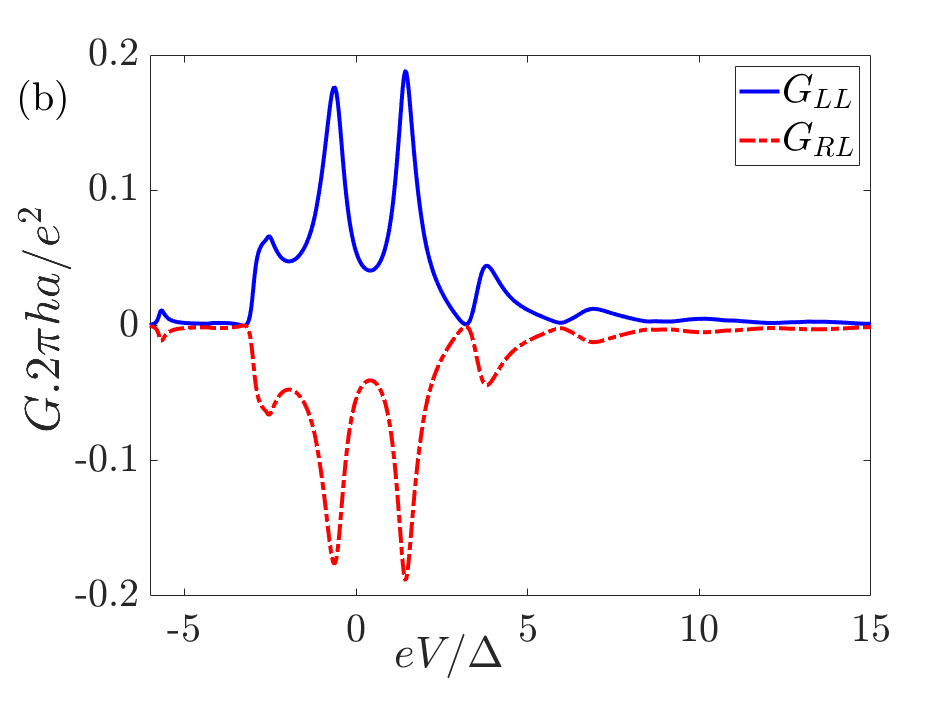}
    
\end{minipage}

\par\medskip

\begin{minipage}[t]{0.48\textwidth}
    \centering
    \includegraphics[width=\linewidth]{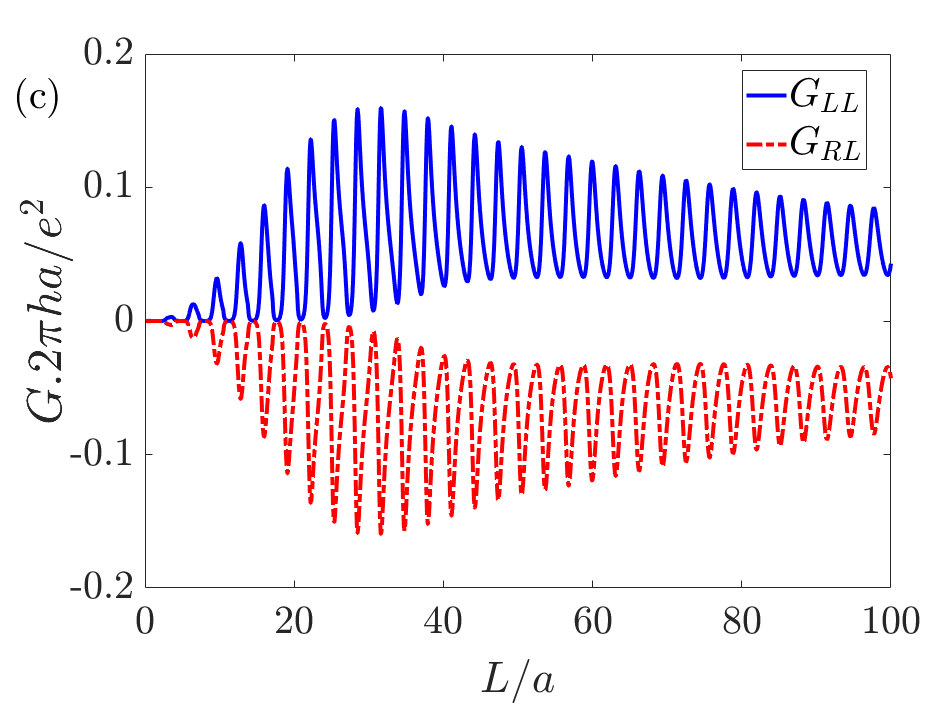}
    
\end{minipage}
\hfill
\begin{minipage}[t]{0.48\textwidth}
    \centering
    \includegraphics[width=\linewidth]{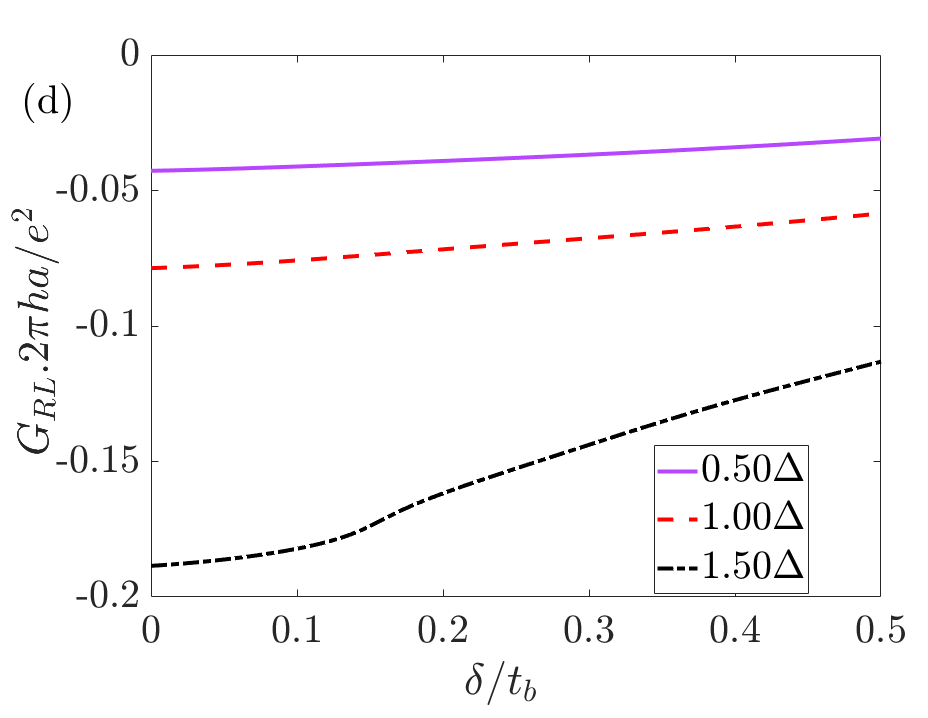}
    
\end{minipage}

    \caption{Local and nonlocal conductivities: (a) versus applied bias in the presence of SOC ($\alpha=0.2t_b$) , $\delta = 0.1t_b$  (b) versus applied bias without SOC; $\delta = 0.1t_b$ (c) versus superconducting length for $\mu_{bl}=-\mu_{br}=0.5t_b$ , $\delta = 0.1t_b$; $\alpha = 0$ and (d) non local conductivity versus $\delta/t_b$ with $\alpha = 0$. The other parameters are $\mu_s=t_b$, $\theta=0$, $b=0.38t_b$, $L=100.6a$, $\Delta=0.02t_b$, and $q_0=0.2\sqrt{\mu_s/(t_s a^2)}$.}
    \label{fig:conductivity}
\end{figure*}

The probabilities of the scattering processes are shown in Fig.~\ref{fig:probabilities_SOC}(a) as a function of the incident electron energy E for normal incidence from the left BMS, including SOC, and in Fig.~\ref{fig:probabilities_SOC}(b) for the case in which SOC is neglected in the BMS. These probability curves exhibit pronounced peaks at specific energies, originating from Fabry--Pérot--type interference driven by multiple back-and-forth reflections within the superconducting region. In the absence of  SOC, the local and nonlocal conductivities exhibit equal magnitudes but with opposite signs. This behavior originates from the suppression of ET and AR, such that  ER and CAR become the only allowed local and nonlocal transport processes, respectively. In the local transport channel, ER involves the reflection of an incident electron back into the same  lead, whereas in the nonlocal channel, CAR converts an incident electron into a hole propagating into the opposite  lead. Since electrons and holes carry opposite charges, their contributions to the local and nonlocal conductivities acquire opposite signs. In contrast, when SOC is finite, both ET and CAR contribute to the transport, along with the other allowed scattering processes. The coexistence of these channels modifies the balance between the local and nonlocal conductivities, such that they are no longer equal in magnitude. In particular, the finite contribution from the ET process can compensate for the negative contribution associated with CAR and may even lead to a positive nonlocal conductivity. Thus, the inclusion of SOC substantially modifies the transport characteristics by enabling multiple scattering channels and altering their relative contributions to the conductivity. 

The peaks observed in the conductivity spectra stem from Fabry--Pérot--type oscillations in the CAR transmission probability, $T_{CAR}$, analogous to the behavior shown in Fig.~\ref{fig:probabilities_SOC}(a),(b).
Figure~\ref{fig:probabilities_SOC}(c) illustrates the CAR probability as a function of $\mu_{0} = \mu_{bl} = -\mu_{br}$, with all other parameters held constant. We observe that the CAR probability generally increases with $\mu_0$. For values of $\mu_0$ near zero, the BMS lacks an available conduction band at zero energy to support incident electrons. However, once $\mu_0$ exceeds a critical threshold, conducting bands emerge on both sides of the SC, yielding a finite CAR probability. As $\mu_0$ is increased further, the wavenumbers of the conducting channels in the BMS leads come closer to the real part of the SC's wavenumber, systematically enhancing the CAR probability. 

In Fig.~\ref{fig:probabilities_SOC}(d), the probability of CAR decreases with increasing SOC strength, as competing transport mechanisms, such as  ET ,ER and AR, become increasingly prominent and suppress the relative contribution of CAR to the overall transport. Beyond a critical SOC strength, ET becomes the dominant transport mechanism, surpassing CAR.

The resonance peaks in the conductivity spectra arise from Fabry--P\'erot--type oscillations in the CAR transmission probability, $T_{CAR}$, consistent with the patterns illustrated in Figures~\ref{fig:conductivity}(a) and (b). As depicted in Figure~\ref{fig:conductivity}(c), both the local and non-local conductivities display oscillatory behavior as a function of the SC length, modulated by an exponentially decaying envelope. Sub-gap excitations cannot propagate freely through the SC; instead, they manifest as evanescent waves, causing the conductivity to decay over a characteristic length scale governed by the superconducting coherence length. The superimposed oscillatory pattern arises from the phase-coherent interference of Bogoliubov--de Gennes quasiparticles, which accumulate a momentum-dependent dynamical phase across the superconducting region despite their evanescent nature.The real and imaginary parts of the superconducting wave vector govern the oscillatory and decaying behavior of the conductivity, respectively. Initially, the conductivity increases with increasing superconducting length due to the enhanced electron-to-hole conversion. However, beyond a characteristic length scale associated with the superconducting coherence length, the conductivity begins to decay as the  propagation of the converted holes to the opposite normal lead becomes increasingly suppressed, thereby supressing the nonlocal transport. 
The separation between consecutive peaks can be understood using the Fabry--P\'erot interference condition: the phase accumulated in one round trip, $2 \,\mathrm{Re}(q_x) L$, must equal an integer multiple of $2\pi$. Accordingly, the peak separation is given by $\Delta L = \pi / \mathrm{Re}(q_{x1/x3}) \approx 3.2\,a$, in good agreement with the value observed in the plot. Here, we have chosen the case of normal incidence since that gives maximum contribution to the conductivity.

Figure~\ref{fig:conductivity}(d) illustrates the nonlocal conductivity as a function of $\delta/t_b$ in the absence of SOC, where $\delta$ characterizes the difference between the effective masses of the spin-up and spin-down electrons. As $\delta$ increases, the nonlocal conductivity decreases in magnitude. This behavior arises because, without SOC, the local and nonlocal conductivities are equal in magnitude but opposite in sign. Furthermore, in the limit of $\alpha=0$, the local conductivity is directly proportional to the wavenumber $k_{e1l}$ [see Eq.~\eqref{eq:Jc} and Eq.~\eqref{eq:conductivity}]. Since $k_{e1l} \propto 1/\sqrt{t_b+\delta}$, an increase in $\delta$ inherently reduces the magnitude of the nonlocal conductivity.

\section{Summary and Conclusion}

In summary, we have theoretically investigated the quantum transport properties of a hybrid device consisting of a central superconducting region coupled to two BMS leads, considering both the presence and absence of SOC in the leads.  In the presence of SOC finite probabilities of AR and ET reduce the relative contribution of CAR . Furthermore, the local and nonlocal conductivities exhibit unequal magnitudes because the current is not exclusively governed by CAR and ER, but also receives contributions from other competing scattering processes. By comparing the transport characteristics of the systems with and without SOC, we demonstrate how SOC modifies the interplay between superconductivity and spin-dependent transport mechanisms, thereby influencing the overall conductance behavior and the efficiency of nonlocal transport in superconducting hybrid systems.
 By exploiting the unique spin-polarized band structure of BMSs, we demonstrated a purely electrical method for tuning CAR without the need for external magnetic fields. This approach eliminates the stray magnetic fields that typically complicate experimental realizations and interfere with nearby components in conventional FM based systems.

Our analysis reveals that tuning the chemical potentials of the two BMS leads to opposite signs ensures that electrons participating in transport on the two sides possess opposite spins at the Fermi energy.In absence of SOC , this specific configuration rigorously suppresses both local AR and ET. Consequently, the transport regime becomes completely dominated by the direct competition between CAR and ER.  In this pure CAR-dominated regime, we found that the non-local conductivity is negative and precisely matches the magnitude of the local conductivity.

Furthermore, we observed pronounced Fabry--Pérot--type resonances in the CAR transmission probability, which arise from multiple back-and-forth reflections within the superconducting region. Both local and non-local conductivities exhibit an oscillatory behavior as a function of the SC length, modulated by an exponentially decaying envelope. This behavior is a direct manifestation of the phase-coherent interference of evanescent Bogoliubov--de Gennes quasiparticles, which acquire a momentum-dependent dynamical phase across the superconducting gap. 
We further observe that weak  SOC leads to only a slight reduction in the  CAR efficiency, indicating that CAR remains a significant transport mechanism in the weak-SOC regime. This suggests that the CAR process exhibits relative robustness against weak SOC, whereas increasing the SOC strength can more substantially modify the interplay between competing scattering mechanisms, thereby influencing the overall conductivity characteristics and the efficiency of nonlocal Cooper-pair splitting in superconducting hybrid structures.

Ultimately, these findings establish that BMS-SC-BMS heterostructures provide a robust pathway for generating and manipulating nonlocal superconducting correlations. This purely electrostatic control significantly simplifies device architecture and highlights a highly efficient, electrically tunable mechanism for scalable superconducting device integration.

\section*{ Data Availability Statement}
Data is available from the authors upon reasonable request.
\begin{acknowledgments}
PN thanks Council of Scientific \& Industrial Research for financial support. AS thanks Science and Engineering Research Board (now Anusandhan National Research Foundation) Core Research grant (CRG/2022/004311) and University of Hyderabad for financial support. 
\end{acknowledgments}
    \bibliography{references.bib}

\end{document}